\documentclass[aps,prx,reprint,amsmath,amssymb,floatfix,citeautoscript,superscriptaddress]{revtex4-1}
\usepackage{cancel}
\usepackage{amsmath}
\usepackage{physics}
\usepackage{graphicx}
\usepackage{amssymb}
\usepackage{amsthm}
\usepackage{bm}
\usepackage{dcolumn}
\usepackage{braket}
\usepackage{longtable}

\usepackage{ragged2e}
\usepackage{txfonts}

\usepackage{color}
\usepackage[usenames,dvipsnames]{xcolor}
\definecolor{myblue}{rgb}{0,0,1}
\usepackage[breaklinks=true,colorlinks=true,linkcolor=myblue,urlcolor=myblue,citecolor=myblue]{hyperref}

\begin{document}

\title{
Linear free energy relationships in electrostatic catalysis
}

\author{Norah M. Hoffmann}
\thanks{These two authors contributed equally}
\affiliation{Department of Chemistry, Columbia University, New York, New York 10027 USA}
\author{Xiao Wang}
\thanks{These two authors contributed equally}
\affiliation{Center for Computational Quantum Physics, Flatiron Institute, New York, New York 10010 USA}
\author{Timothy C. Berkelbach}
\email{tim.berkelbach@gmail.com}
\affiliation{Department of Chemistry, Columbia University, New York, New York 10027 USA}
\affiliation{Center for Computational Quantum Physics, Flatiron Institute, New York, New York 10010 USA}

\begin{abstract}
The use of electric fields to modify chemical reactions is a promising, emerging
technique in catalysis. However, there exist few guiding principles, and
rational design requires assumptions about the transition state or explicit
atomistic calculations. Here, we present a linear free energy relationship,
familiar in other areas of physical organic chemistry, that microscopically
relates field-induced changes in the activation energy to those in the reaction
energy, connecting kinetic and thermodynamic behaviors. We verify our theory
using first-principles electronic structure calculations of a symmetric
S$_\mathrm{N}$2 reaction and the dehalogenation of an aryl halide on gold
surfaces and observe hallmarks of linear free energy relationships,such as the
shifting to early and late transition states. We also report and explain a
counterintuitive case, where the constant of proportionality relating the
activation and reaction energies is negative, such that stabilizing the product
increases the activation energy, i.e., opposite of the Bell-Evans-Polanyi
principle.
\end{abstract}

\maketitle

Recent years have seen an increase in the study of electric field effects on
chemical
reactions~\cite{shaik2016oriented,che2018elucidating,ciampi2018harnessing,
foroutan2014potential, lau2017electrostatic, gorin2012electric,
gorin2013interfacial,welborn2018computational,shaik2020electric,joy2020oriented},
motivated in part by the large electric fields that have been implicated in the
catalytic performance of biological
enzymes~\cite{kamerlin2010ketosteroid,fried2014extreme,
warshel2006electrostatic, welborn2019fluctuations}.  Chemical reactions, whose
transitions states have large dipole moments are natural targets for
electrostatic catalysis, because an external electric field will modify the
activation energy and thus the reaction kinetics.  However, as a rational design
principle, this requires knowledge of the transition state geometry and
electronic structure, which is nontrivial to obtain. In contrast, the reactant
and product, which define the reaction thermodynamics, are much more easily
characterized.  Here, we derive and computationally test a linear free energy
relationship, which relates field-induced changes in the activation energy to
those in the reaction energy, providing a powerful new framework for design and
characterization of electrostatic catalysis.

In the presence of a uniform electric field $F$, the change in free energy $G$ of 
a molecular system is approximately given by
$\Delta G = - \mu F$, where $\mu$ is the component of the molecular dipole moment in the direction of the
field.
We consider a chemical reaction progressing through a reactant (R), transition state (TS), 
and product (P) geometry. 
Assuming that the field does not qualitatively alter the reaction pathway, the activation energy 
$\Delta G^\ddagger = G_\mathrm{TS}-\mathrm{G}_\mathrm{R}$ 
and reaction energy $\Delta G_\mathrm{rxn} = G_\mathrm{P}-G_\mathrm{R}$ are modified according to
\begin{align}
\label{eq:ddg_act}
\Delta\Delta G^\ddagger 
    &= - (\mu_\mathrm{TS}-\mu_\mathrm{R})F \equiv -\Delta\mu^\ddagger F\\
\label{eq:ddg_rxn}
\Delta\Delta G_\mathrm{rxn} 
    &= - (\mu_\mathrm{P}-\mu_\mathrm{R})F \equiv -\Delta\mu_\mathrm{rxn}F
\end{align}
Combining Eqs.~(\ref{eq:ddg_act}) and (\ref{eq:ddg_rxn}) gives the linear free energy relationship (LFER)
\begin{subequations}
\label{eq:lfer}
\begin{align}
\Delta\Delta G^\ddagger &= m \Delta\Delta G_\mathrm{rxn} \\
m &= \frac{\Delta\mu^\ddagger}{\Delta\mu_\mathrm{rxn}}
\end{align}
\end{subequations}
Assuming an Arrhenius form to the reaction rate with a prefactor that is independent of the
field yields the relationship $\log k_F/k_0 = m\log K_F/K_0$, 
where $k_F$, $K_F$ ($k_0$, $K_0$) are the reaction rate constant and equilibrium constant
in the presence (absence) of a field.
Therefore, the LFER~(\ref{eq:lfer}) quantitatively relates the kinetics and thermodynamics
of chemical reactions in electric fields. 
 
Unlike most LFERs that are empirical, the LFER~(\ref{eq:lfer}) provides
a microscopic expression for the constant of proportionality or slope, 
$m = \Delta\mu^\ddagger/\Delta\mu_\mathrm{rxn}$,
whose magnitude determines the sensitivity of the activation energy to changes in the reaction energy.
In particular, it encodes the common wisdom that reactions, whose transition state 
has a large dipole moment are especially susceptible to catalysis by an electric field,
because in that case $|\Delta \mu^\ddagger| \gg 0$. However, the slope
$m$ can have positive or negative sign. The
case with positive slope signals that an electric field that stabilizes
the product with respect to the reactant will also lower the activation energy;
this behavior is conventional and forms the basis of the Bell-Evans-Polanyi principle~\cite{van2009reactivity}. 
The case with negative slope is unconventional and predicts the opposite
behavior: applying an electric field that lowers the reaction energy will
\textit{increase} the activation energy.

\begin{figure}[t]
\includegraphics[width=1.0\columnwidth]{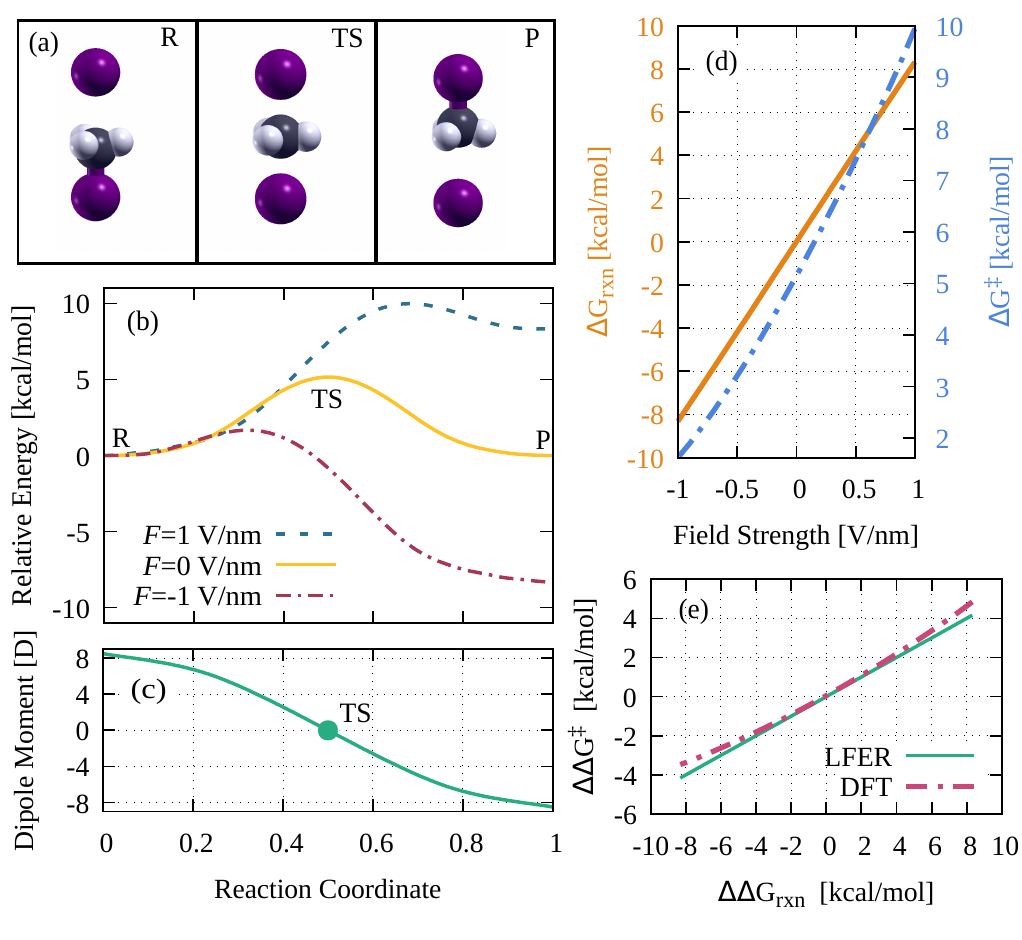}
\caption{Symmetric S$_\mathrm{N}$2 reaction of I$^{-}$ with methyl iodide. 
(a) Geometries of reactant (R), transition state (TS), and product (P) in the absence
of a field.
(b) Reaction energy in the absence and presence of a field of strengths indicated.
(c) Dipole moment evolution in the absence of a field.
(d) Reaction energy (orange, left axis) and activation energy (blue, right axis) as
a function of field strength.
(e) Comparison of the LFER prediction to the DFT simulation results.
}
\label{fig:sn2}
\end{figure}

To elucidate the theory, we first consider a simple, symmetric S$_\mathrm{N}$2 reaction of
I$^{-}$ with methyl iodide [Fig.~\ref{fig:sn2}]. 
Using density functional theory (DFT) with the B3LYP functional,
we calculated the minimum energy reaction pathway
in the absence and presence of an external
electric field directed along the I-C-I bond axis [Fig.~\ref{fig:sn2}(b)]. 
First we observe that, over the range of field
strengths $-1$ to $1$~V/nm, the activation
energy can be tuned by over 8~kcal/mol, corresponding to a factor of $10^6$ in the reaction
rate at room temperature and highlighting the potential power of electrostatic catalysis.
More important for the present work, we observe that 
the reaction energy and activation energy are found
to vary roughly linearly with the field strength [Fig.~\ref{fig:sn2}(d)], in agreement with Eqs.~(\ref{eq:ddg_act}) and
(\ref{eq:ddg_rxn}) (deviations from linearity at larger field strengths are due
to molecular polarizabilities and changes in the geometries). 
Therefore, the LFER~(\ref{eq:lfer}) holds to an excellent approximation, with
$\Delta\Delta G^\ddagger \approx (+1/2) \Delta\Delta G_\mathrm{rxn}$ [Fig.~\ref{fig:sn2}(e)].

The observed slope
$m\approx+1/2$ is readily explained by an analysis of the evolution of the dipole moment over the
course of the reaction in the \textit{absence} of a field [Fig.~\ref{fig:sn2}(c)]. 
Importantly, the dipole moment decreases monotonically,
from $\mu=8.4$~D (R) to 0.0~D (TS) to $-8.4$~D (P).
Therefore, the dipole moment differences $\Delta \mu^\ddagger = -8.4$~D and 
$\Delta\mu_\mathrm{rxn} = -16.8$~D have the same (negative) sign, and their ratio, the slope $m$, is
positive. Moreover, the magnitude of the slope is explained by the ratio
$m=\Delta \mu^\ddagger/\Delta \mu_\mathrm{rxn} \approx (-8.4)/(-16.8)=+1/2$, in agreement
with the computational observation. 

The field-dependent reaction profiles in Fig.~\ref{fig:sn2}(b) also show the behavior 
of ``early'' and ``late'' TSs, consistent with Hammond's postulate.
For example, in the presence of a field with $F=-1$~V/nm, the reaction is more exothermic, has a lower activation
energy, and has an early TS. The change in the TS
geometry can be quantified: the bond length between the attacking iodine and the
carbon is found to be 2.60~\AA, 2.72~\AA, and 2.85~\AA\ and the angle made 
by the the attacking iodine,
carbon, and hydrogen is found to be 94$^\circ$, 90$^\circ$, and 86$^\circ$,
in the early, middle, and late TS geometries, respectively.
We conclude that simple reactions influenced by electric fields obey the LFER~(\ref{eq:lfer}) to a good
approximation and exhibit many of its hallmarks known from synthetic organic chemistry. 

Next, we test our theory on a more complicated example from heterogeneous catalysis.
Specifically, we computationally study the dehalogenation of an iodobenzene molecule
on a gold surface. This reaction is a key step in the metal-catalyzed Ullmann coupling
reaction of aryl halides~\cite{bjork2013mechanisms}. 
Participation of the metal is crucial, because the aryl halide bond dissociation energy
is about $+70$~kcal/mol in the gas phase.
All calculations are performed with periodic DFT with the optB86b-vdW functional, whose accuracy
has been established for noncovalent interactions~\cite{klimevs2011van}, and we
identify minimum-energy reaction pathways using the climbing-image nudged
elastic band approach.

We apply an electric field in the direction perpendicular to the gold surface with both
positive and negative sign. This electric field may be internal, due to the electrostatics
of a solid-liquid interface or local field effects. The field may
also be external and applied controllably using the gold tip of a scanning tunneling 
microscope (STM), which has recently been used in a break-junction arrangement to 
experimentally demonstrate electrostatic catalysis~\cite{meir2010oriented,zang2019directing}.
Again, we consider a maximum field strength of 1~V/nm, which is appropriate for
these experimental realizations.
We emphasize that quantitative properties of the electric field will depend on 
atomistic details of the surface and solvent as well as their dynamics, and 
our study is only qualitative. 

\begin{figure}[b]
\includegraphics[width=1.0\columnwidth]{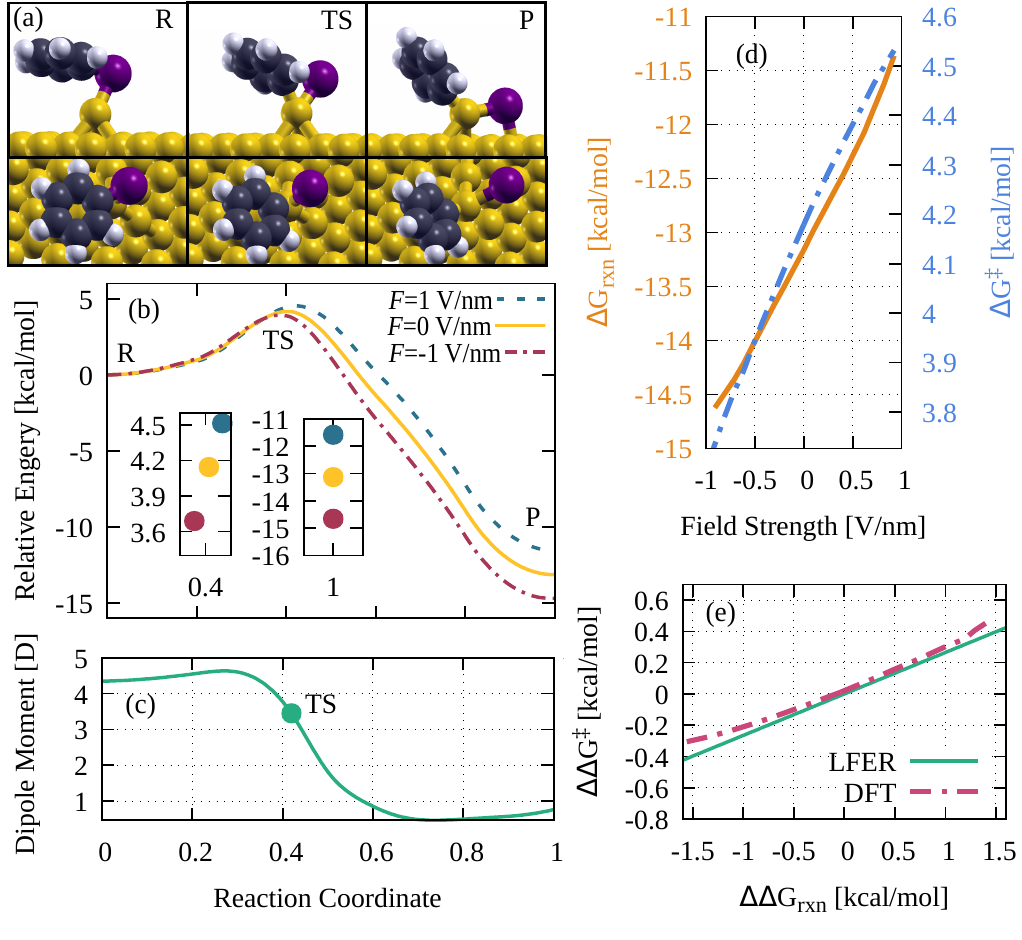}
\caption{Same as Fig.~\ref{fig:sn2}, for the dehalogenation of iodobenzene on 
a (111) gold surface with a single adatom. In (a), geometries are shown from the side and from above.
In (b), insets show a close-up of the energies and locations of the transition state and product.
In (c), the plotted dipole moment is only the component in the $z$ direction, which is the direction of the
applied field.
}
\label{fig:rough}
\end{figure}

To model the surface roughness, we first study the dehalogenation reaction
on a gold (111) surface with an additional gold adatom at a hollow site (Fig.~\ref{fig:rough}). 
In the absence of a field, we find a 
reaction energy of $\Delta G_\mathrm{rxn} = -13.4$~kcal/mol and an
activation energy of $\Delta G^\ddagger = +4.1$~kcal/mol. Notably, this activation energy
is significantly less than the one previously calculated for the same reaction on a clean gold
surface, i.e., without an adatom, which was +16.9~kcal/mol~\cite{bjork2013mechanisms}. This large
reduction in the activation energy agrees with the common wisdom that rough
metal surfaces are better catalysts than clean metal surfaces.

Assuming that the reaction pathway does not change in the presence of the field,
we recalculate the electronic energy in the presence of an external electric
field in the range $F=-1$~V/nm to $F=+1$~V/nm [Fig.~\ref{fig:rough}(b)], where a positive field
points away from the surface, i.e., in the positive $z$-direction.
(Preliminary testing indicates that reoptimizing the geometries in the presence of the field does
not significantly alter our findings over the range of fields considered here.)
We find that application of an electric
field indeed catalyzes the reaction, albeit only slightly: 
a field of $F=+1$~V/nm lowers the activation energy by
0.45~kcal/mol, which would increase the reaction rate by a factor of two
at room temperature. Reversing the direction of
the field raises the activation energy by a similar amount and thus supresses the reaction.
More importantly, the reaction energy and activation energy are again observed to
be roughly linear in the field strength [Fig.~\ref{fig:rough}(d)] and the TS
shifts to an earlier one as the reaction becomes more favorable [inset of Fig.~\ref{fig:rough}(b)]. 
Similar to the S$_\mathrm{N}$2 behavior,
the LFER~(\ref{eq:lfer}) holds approximately with a positive slope less than one,
$\Delta\Delta G^\ddagger \approx (+0.3)\Delta\Delta G_\mathrm{rxn}$ [Fig.~\ref{fig:rough}(e)]. 

The sign and magnitude of the slope can again be explained by analyzing the
evolution of the dipole moment (in the direction of the field) over the course of the
reaction in the absence of the field
from $\mu=4.4$~D (R) to 3.4~D (TS) to 0.75~D (P)
 [Fig.~\ref{fig:rough}(c)]. 
Therefore, the dipole moment differences $\Delta \mu^\ddagger = -1.0$~D
and $\Delta \mu_\mathrm{rxn}=-3.6$~D have the same (negative) sign, and their ratio, the slope $m$,
is positive. Moreover, the magnitude of the slope is explained by the ratio 
$\Delta \mu^\ddagger/\Delta \mu_\mathrm{rxn} \approx 0.3$, which is why the change in the activation
energy is about three times smaller than the change in the reaction energy.

Unlike for the simple S$_\mathrm{N}$2 reaction, the behavior of the dipole moment
for this surface-mediated dehalogenation is not easily rationalized, because the
largest changes to the electronic structure occur in the direction parallel to
the surface, i.e., perpendicular to the field.
However, analysis of our DFT calculations in terms of electron densities and L\"owdin charges (not shown)
can provide some insight into the evolution of the dipole moment in the $z$ direction. 
In the chemisorbed reactant, there is significant electron transfer from the molecule to the gold,
establishing a dipole moment in the positive $z$ direction. 
The reaction occurs via an oxidative addition-like pathway, whereby electron transfer
from the undercoordinated gold adatom onto the aryl halide facilitates
bond breaking and slightly reduces the dipole moment of the TS. In the product geometry, the
aryl ring has a greater electron density and thus a less positive total charge, which we attribute
to improved resonance with the gold orbitals. Overall, the charge separation between molecule and gold
is smallest in the product, consistent with its small dipole moment.

In our opinion, the previous two computational studies confirm the most important result of
the present work, i.e., that a LFER holds for electrostatic catalysis and that,
\textit{if its slope is positive}, changes to the activation energy are
positively proportional to changes in the reaction energy.  For the endeavor of
electrostatic catalysis, this finding shifts focus from the activation
energy---and its associated challenge of transition state design---to the much
more manageable reaction energy. Alternatively, an experimental measurement of the slope $m$
appearing in the LFER~(\ref{eq:lfer})
can be used to infer details about the reaction pathway, including its TS and associated dipole moment
differences. We expect the positivity of the slope to hold
in many, but not all, reactions. Specifically, a sufficient, but not necessary
condition for the positivity of the slope is that the dipole moment evolution is
monotonic (increasing or decreasing) over the course of the reaction, which is a
common and intuitive behavior. However, many important reactions have a
transition-state dipole moment that is larger than that of the reactant and
product. In this case, the slope may be negative, which we demonstrate now with
a closely related computational example.

\begin{figure}[b]
\includegraphics[width=1.0\columnwidth]{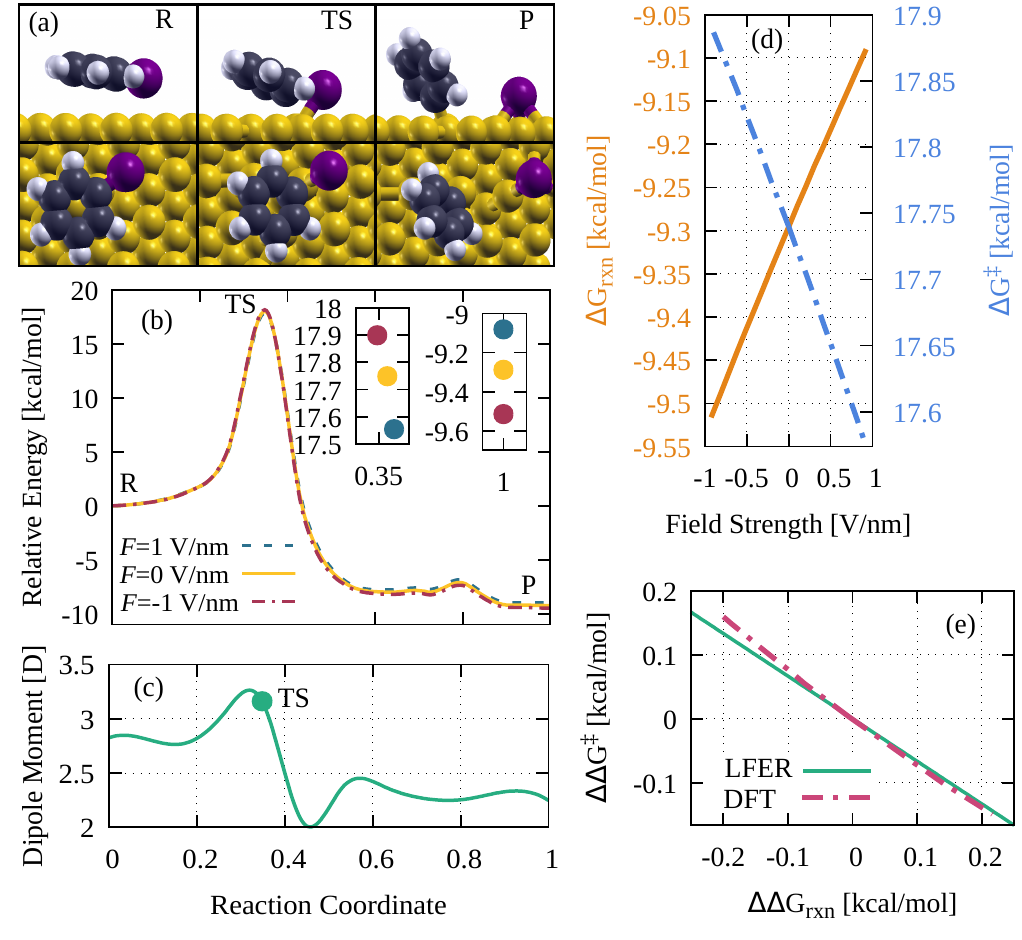}
\caption{Same as Fig.~\ref{fig:rough}, for the dehalogenation of iodobenzene on 
a clean (111) gold surface.}
\label{fig:clean}
\end{figure}

We study the same dehalogenation reaction on a clean gold (111) surface, i.e., without an adatom
[Fig.~\ref{fig:clean}]. In this case, oxidative addition by gold is sterically hindered and the
activation energy is signficantly higher, $\Delta\Delta G^\ddagger = 17.74$~kcal/mol. 
The activation energy and reaction energy are again linear in the strength of the field [Fig.~\ref{fig:clean}(d)]
and can be weakly tuned by about 
0.3~kcal/mol and 0.4~kcal/mol, respectively. 
More interestingly, the LFER~(\ref{eq:lfer}) holds, but with a \textit{negative} slope [Fig.~\ref{fig:clean}(e)].
In other words, the field orientation that lowers the reaction energy actually \textit{raises} the activation energy and
shifts the TS to an earlier one~[Fig.~\ref{fig:clean}(b)]. This negative slope is explained by the
dipole moment, which is nonmonotonic over the course of the reaction, evolving from 2.8~D~(R) to 3.2~D~(TS) to 2.2~D~(P),
i.e., with $\Delta \mu^\ddagger = +0.4~$D and $\Delta\mu_{\mathrm{rxn}} = -0.6$~D~[Fig.~\ref{fig:clean}(c)]. 
Thus the LFER~(\ref{eq:lfer}) predicts a slope of $m \approx (+0.4)/(-0.6) = -(2/3)$, in good agreement
with the computational result.

Again, analysis of our DFT calculations can rationalize the behavior of the
dipole moment. We find that the charge assignments of the reactant and product
are qualitatively similar to those in the presence of the adatom. However, we
find that in the TS, rather than accepting electron density from the gold, the
aryl iodide molecule donates additional electron density to the gold, increasing
the extent of charge separation and thus increasing the magnitude of the dipole
moment. This example shows how the analysis of reactions in terms of the LFER~(\ref{eq:lfer})
can be used to uncover unusual mechanistic details of complex chemical reactions.

The theory developed here is general and can be readily applied in other situations. For example, the
electric field need not be externally applied, and could be self-induced, as was
recently discussed for a molecule in a plasmonic
nanocavity~\cite{climent2019plasmonic}. Specifically,
a molecule with dipole moment $\mu$ sandwiched between two metals will induce a
proportional reaction field $F\propto \mu$,
which lowers the molecular energy by $\Delta G \propto \mu^2$. A LFER naturally
follows with the form 
$\Delta\Delta G^\ddagger = m' \Delta\Delta G_\mathrm{rxn}$, where now the slope
is given by the difference in the square of the dipole moments,
\begin{equation}
m' = \frac{\mu^2_\mathrm{TS}-\mu^2_\mathrm{R}}{\mu^2_\mathrm{P}-\mu^2_\mathrm{R}}.
\end{equation}
Therefore, we expect the work presented here, which provided a general, but
microscopic relationship between the kinetics and thermodynamics of chemical
reactions in electric fields to have impact throughout the field of
electrostatic catalysis, similar to how scaling relations have revolutionized
heterogeneous catalysis~\cite{Norskov2002,Michaelides2003,Bligaard2004}.

\vspace{1em}

\section*{Acknowledgments}
This work was supported primarily by NSF CHE-2023568 CCI Phase I: Center for
Chemistry with Electric Fields (N.M.H. and T.C.B.). We thank Ilana Stone, Rachel Starr, Xavier Roy,
and Latha Venkataraman for useful discussions. We acknowledge computing
resources from Columbia University's Shared Research Computing Facility project,
which is supported by NIH Research Facility Improvement Grant 1G20RR030893-01,
and associated funds from the New York State Empire State Development, Division
of Science Technology and Innovation (NYSTAR) Contract C090171, both awarded
April 15, 2010. 
The Flatiron Institute is a division of the Simons Foundation.

\section*{Computational Details}

All DFT calculations for the S$_\mathrm{N}$2 reaction were performed using the Orca
\cite{neese2012wiley} quantum chemistry package with the B3LYP
functional~\cite{becke1988density, becke1993new} and the def2-TZVP
basis set~\cite{schafer1994fully, weigend2005balanced}. 
The climbing-image nudged elastic band method
\cite{mills1994quantum, mills1995reversible, berne1998classical,
henkelman2000climbing, henkelman2000improved} followed by the eigenvalue-based
TS searching algorithm \cite{banerjee1985search, baker1986algorithm} (Orca keyword
``NEB-TS'') was used to find the minimum energy path and the TS structure. 
For each applied field, the reaction path geometries were fully optimized.

All periodic DFT calculations for the dehalogenation reaction on a gold 
surface were performed using the Quantum
Espresso package~\cite{giannozzi2009quantum}. Following Ref.~\onlinecite{bjork2013mechanisms}, 
we applied the optB86b-vdW functional~\cite{klimevs2011van},
which provides an affordable description of dispersion interactions, yielding
results comparable to those obtained by the random phase approximation
\cite{bjork2013mechanisms,klimevs2011van}. Core electrons were treated using the projector 
augmented wave method~\cite{blochl1994projector, blochl1994improved}. We used
a 500~eV kinetic energy cutoff and a $2\times 2\times 1$ sampling of the Brillouin zone.
Gold surfaces were modeled using $5\times 5$ slabs, 15~$\AA$ of vacuum, and a lattice
constant of 4.140~$\AA$.
The clean gold (111) surface was modeled by a four-layer slab, and the two uppermost layers
were allowed to relax during geometry optimizations. The system with an
extra adatom was modeled by a three-layer slab, and the adatom and uppermost layer were
allowed to relax. The transition state calculations were performed with the climbing image
nudged elastic band approach, using 12 images.
The electric field was applied with a sawtooth potential~\cite{giannozzi2009quantum}
and all calculations were dipole-corrected~\cite{bengtsson1999dipole} using a
dipole length of 0.89~$\AA$ in the direction of the field.

\bibliographystyle{unsrt}

\begin{thebibliography}{}

\end{thebibliography}


\begin{thebibliography}{10}

\bibitem{shaik2016oriented}
Sason Shaik, Debasish Mandal, and Rajeev Ramanan.
\newblock Oriented electric fields as future smart reagents in chemistry.
\newblock {\em Nat. Chem.}, 8(12):1091--1098, 2016.

\bibitem{che2018elucidating}
Fanglin Che, Jake~T Gray, Su~Ha, Norbert Kruse, Susannah~L Scott, and
  Jean-Sabin McEwen.
\newblock Elucidating the roles of electric fields in catalysis: A perspective.
\newblock {\em ACS Catal.}, 8(6):5153--5174, 2018.

\bibitem{ciampi2018harnessing}
Simone Ciampi, Nadim Darwish, Heather~M Aitken, Ismael D{\'\i}ez-P{\'e}rez, and
  Michelle~L Coote.
\newblock Harnessing electrostatic catalysis in single molecule,
  electrochemical and chemical systems: a rapidly growing experimental tool
  box.
\newblock {\em Chem. Soc. Rev.}, 47(14):5146--5164, 2018.

\bibitem{foroutan2014potential}
Cina Foroutan-Nejad and Radek Marek.
\newblock Potential energy surface and binding energy in the presence of an
  external electric field: modulation of anion--$\pi$ interactions for
  graphene-based receptors.
\newblock {\em Phys. Chem. Chem. Phys.}, 16(6):2508--2514, 2014.

\bibitem{lau2017electrostatic}
Vivian~M Lau, William~C Pfalzgraff, Thomas~E Markland, and Matthew~W Kanan.
\newblock Electrostatic control of regioselectivity in au (i)-catalyzed
  hydroarylation.
\newblock {\em J. Am. Chem. Soc.}, 139(11):4035--4041, 2017.

\bibitem{gorin2012electric}
Craig~F Gorin, Eugene~S Beh, and Matthew~W Kanan.
\newblock An electric field--induced change in the selectivity of a metal
  oxide--catalyzed epoxide rearrangement.
\newblock {\em J. Am. Chem. Soc.}, 134(1):186--189, 2012.

\bibitem{gorin2013interfacial}
Craig~F Gorin, Eugene~S Beh, Quan~M Bui, Graham~R Dick, and Matthew~W Kanan.
\newblock Interfacial electric field effects on a carbene reaction catalyzed by
  rh porphyrins.
\newblock {\em J. Am. Chem. Soc.}, 135(30):11257--11265, 2013.

\bibitem{welborn2018computational}
Valerie~Vaissier Welborn, Luis~Ruiz Pestana, and Teresa Head-Gordon.
\newblock Computational optimization of electric fields for better catalysis
  design.
\newblock {\em Nat. Catal.}, 1(9):649--655, 2018.

\bibitem{shaik2020electric}
Sason Shaik, David Danovich, Jyothish Joy, Zhanfeng Wang, and Thijs Stuyver.
\newblock Electric-field mediated chemistry: Uncovering and exploiting the
  potential of (oriented) electric fields to exert chemical catalysis and
  reaction control.
\newblock {\em J. Am. Chem. Soc.}, 142(29):12551--12562, 2020.

\bibitem{joy2020oriented}
Jyothish Joy, Thijs Stuyver, and Sason Shaik.
\newblock Oriented external electric fields and ionic additives elicit
  catalysis and mechanistic crossover in oxidative addition reactions.
\newblock {\em J. Am. Chem. Soc.}, 142(8):3836--3850, 2020.

\bibitem{kamerlin2010ketosteroid}
Shina~CL Kamerlin, Pankaz~K Sharma, Zhen~T Chu, and Arieh Warshel.
\newblock Ketosteroid isomerase provides further support for the idea that
  enzymes work by electrostatic preorganization.
\newblock {\em Proc. Natl. Acad. Sci.}, 107(9):4075--4080, 2010.

\bibitem{fried2014extreme}
Stephen~D Fried, Sayan Bagchi, and Steven~G Boxer.
\newblock Extreme electric fields power catalysis in the active site of
  ketosteroid isomerase.
\newblock {\em Science}, 346(6216):1510--1514, 2014.

\bibitem{warshel2006electrostatic}
Arieh Warshel, Pankaz~K Sharma, Mitsunori Kato, Yun Xiang, Hanbin Liu, and
  Mats~HM Olsson.
\newblock Electrostatic basis for enzyme catalysis.
\newblock {\em Chem. Rev.}, 106(8):3210--3235, 2006.

\bibitem{welborn2019fluctuations}
Valerie~Vaissier Welborn and Teresa Head-Gordon.
\newblock Fluctuations of electric fields in the active site of the enzyme
  ketosteroid isomerase.
\newblock {\em J. Am. Chem. Soc.}, 141(32):12487--12492, 2019.

\bibitem{van2009reactivity}
Rutger~A Van~Santen, Matthew Neurock, and Sharan~G Shetty.
\newblock Reactivity theory of transition-metal surfaces: a br{\o}nsted- evans-
  polanyi linear activation energy- free-energy analysis.
\newblock {\em Chem. Rev.}, 110(4):2005--2048, 2009.

\bibitem{bjork2013mechanisms}
Jonas Bjork, Felix Hanke, and Sven Stafstrom.
\newblock Mechanisms of halogen-based covalent self-assembly on metal surfaces.
\newblock {\em J. Am. Chem. Soc.}, 135(15):5768--5775, 2013.

\bibitem{klimevs2011van}
Ji{\v{r}}{\'\i} Klime{\v{s}}, David~R Bowler, and Angelos Michaelides.
\newblock Van der waals density functionals applied to solids.
\newblock {\em Phys. Rev. B}, 83(19):195131, 2011.

\bibitem{meir2010oriented}
Rinat Meir, Hui Chen, Wenzhen Lai, and Sason Shaik.
\newblock Oriented electric fields accelerate diels--alder reactions and
  control the endo/exo selectivity.
\newblock {\em Chem. Phys. Chem.}, 11(1):301--310, 2010.

\bibitem{zang2019directing}
Yaping Zang, Qi~Zou, Tianren Fu, Fay Ng, Brandon Fowler, Jingjing Yang, Hexing
  Li, Michael~L Steigerwald, Colin Nuckolls, and Latha Venkataraman.
\newblock Directing isomerization reactions of cumulenes with electric fields.
\newblock {\em Nat. Commun.}, 10(1):1--7, 2019.

\bibitem{climent2019plasmonic}
Cl{\`a}udia Climent, Javier Galego, Francisco~J Garcia-Vidal, and Johannes
  Feist.
\newblock Plasmonic nanocavities enable self-induced electrostatic catalysis.
\newblock {\em Angew. Chem. Int. Ed.}, 58(26):8698--8702, 2019.

\bibitem{Norskov2002}
J.K. N{\o}rskov, T.~Bligaard, A.~Logadottir, S.~Bahn, L.B. Hansen,
  M.~Bollinger, H.~Bengaard, B.~Hammer, Z.~Sljivancanin, M.~Mavrikakis, Y.~Xu,
  S.~Dahl, and C.J.H. Jacobsen.
\newblock Universality in heterogeneous catalysis.
\newblock {\em J. Catal.}, 209(2):275--278, jul 2002.

\bibitem{Michaelides2003}
Angelos Michaelides, Z.-P. Liu, C.~J. Zhang, Ali Alavi, David~A. King, and
  P.~Hu.
\newblock Identification of general linear relationships between activation
  energies and enthalpy changes for dissociation reactions at surfaces.
\newblock {\em J. Am. Chem. Soc.}, 125(13):3704--3705, mar 2003.

\bibitem{Bligaard2004}
T.~Bligaard, J.K. N{\o}rskov, S.~Dahl, J.~Matthiesen, C.H. Christensen, and
  J.~Sehested.
\newblock The br{\o}nsted{\textendash}evans{\textendash}polanyi relation and
  the volcano curve in heterogeneous catalysis.
\newblock {\em J. Catal.}, 224(1):206--217, may 2004.

\bibitem{neese2012wiley}
Frank Neese.
\newblock Software update: the orca program system, version 4.0.
\newblock {\em WIREs Comput. Mol. Sci.}, 8(1):e1327, 2018.

\bibitem{becke1988density}
Axel~D Becke.
\newblock Density-functional exchange-energy approximation with correct
  asymptotic behavior.
\newblock {\em Phys. Rev. A}, 38(6):3098, 1988.

\bibitem{becke1993new}
Axel~D Becke.
\newblock A new mixing of hartree--fock and local density-functional theories.
\newblock {\em J. Chem. Phys.}, 98(2):1372--1377, 1993.

\bibitem{schafer1994fully}
Ansgar Sch{\"a}fer, Christian Huber, and Reinhart Ahlrichs.
\newblock Fully optimized contracted gaussian basis sets of triple zeta valence
  quality for atoms li to kr.
\newblock {\em J. Chem. Phys.}, 100(8):5829--5835, 1994.

\bibitem{weigend2005balanced}
Florian Weigend and Reinhart Ahlrichs.
\newblock Balanced basis sets of split valence, triple zeta valence and
  quadruple zeta valence quality for h to rn: Design and assessment of
  accuracy.
\newblock {\em Phys. Chem. Chem. Phys.}, 7(18):3297--3305, 2005.

\bibitem{mills1994quantum}
Greg Mills and Hannes J{\'o}nsson.
\newblock Quantum and thermal effects in h 2 dissociative adsorption:
  Evaluation of free energy barriers in multidimensional quantum systems.
\newblock {\em Phys. Rev. Lett}, 72(7):1124, 1994.

\bibitem{mills1995reversible}
Gregory Mills, Hannes J{\'o}nsson, and Gregory~K Schenter.
\newblock Reversible work transition state theory: application to dissociative
  adsorption of hydrogen.
\newblock {\em Surf. Sci.}, 324(2-3):305--337, 1995.

\bibitem{berne1998classical}
Bruce~J Berne, Giovanni Ciccotti, and David~F Coker.
\newblock {\em Classical and quantum dynamics in condensed phase simulations:
  Proceedings of the International School of Physics}.
\newblock World Scientific, 1998.

\bibitem{henkelman2000climbing}
Graeme Henkelman, Blas~P Uberuaga, and Hannes J{\'o}nsson.
\newblock A climbing image nudged elastic band method for finding saddle points
  and minimum energy paths.
\newblock {\em J. Chem. Phys.}, 113(22):9901--9904, 2000.

\bibitem{henkelman2000improved}
Graeme Henkelman and Hannes J{\'o}nsson.
\newblock Improved tangent estimate in the nudged elastic band method for
  finding minimum energy paths and saddle points.
\newblock {\em J. Chem. Phys.}, 113(22):9978--9985, 2000.

\bibitem{banerjee1985search}
Ajit Banerjee, Noah Adams, Jack Simons, and Ron Shepard.
\newblock Search for stationary points on surfaces.
\newblock {\em J. Phys. Chem.}, 89(1):52--57, 1985.

\bibitem{baker1986algorithm}
Jon Baker.
\newblock An algorithm for the location of transition states.
\newblock {\em J. Comput. Chem.}, 7(4):385--395, 1986.

\bibitem{giannozzi2009quantum}
Paolo Giannozzi, Stefano Baroni, Nicola Bonini, Matteo Calandra, Roberto Car,
  Carlo Cavazzoni, Davide Ceresoli, Guido~L Chiarotti, Matteo Cococcioni,
  Ismaila Dabo, et~al.
\newblock Quantum espresso: a modular and open-source software project for
  quantum simulations of materials.
\newblock {\em J. Condens. Matter Phys.}, 21(39):395502, 2009.

\bibitem{blochl1994projector}
Peter~E Bl{\"o}chl.
\newblock Projector augmented-wave method.
\newblock {\em Phys. Rev. B}, 50(24):17953, 1994.

\bibitem{blochl1994improved}
Peter~E Bl{\"o}chl, Ove Jepsen, and Ole~Krogh Andersen.
\newblock Improved tetrahedron method for brillouin-zone integrations.
\newblock {\em Phys. Rev. B}, 49(23):16223, 1994.

\bibitem{bengtsson1999dipole}
Lennart Bengtsson.
\newblock Dipole correction for surface supercell calculations.
\newblock {\em Phys. Rev. B}, 59(19):12301, 1999.

\end{thebibliography}

\end{document}